\documentclass[useAMS,usenatbib,a4]{mn2e}
\usepackage{times} 
\usepackage{graphicx} 
\usepackage{amsmath}
\newif\ifAMStwofonts 
\AMStwofontstrue

\newcommand{\simlt}{\lower.5ex\hbox{$\; \buildrel < \over \sim \;$}} 
\newcommand{\simgt}{\lower.5ex\hbox{$\; \buildrel > \over \sim \;$}} 
\newcommand{\be}{\begin{equation}} 
\newcommand{\ba}{\begin{eqnarray}}
\newcommand{\ee}{\end{equation}} 
\newcommand{\ea}{\end{eqnarray}}
 
\newcommand{\aFe}{${\boldsymbol [}{\boldsymbol \alpha}/$Fe${\boldsymbol ]}$}
\title 
{The link between the Star Formation History and \aFe\ }

\author[de la Rosa et al.] 
{Ignacio G. de la Rosa$^{1,2,3}$, Francesco La Barbera$^4$,
Ignacio Ferreras$^5$, 
\newauthor Reinaldo R. de Carvalho$^6$\\ 
$^1$ Instituto de Astrof\'\i
sica de Canarias, C/ V\'\i a L\'a ctea s/n, E-38200 La Laguna, Tenerife, Spain\\ 
$^2$ Departamento de Astrof\'\i sica, Universidad de La Laguna, E-38205 La Laguna, Tenerife,
Spain \\ 
$^3$ Department of Physics and Astronomy, University College London, Gower Street,
London, WC1E 6BT \\ 
$^4$ INAF - Osservatorio Astronomico di Capodimonte, Napoli, Italy \\
$^5$ Mullard Space Science Laboratory, University College London,  Holmbury St Mary, Dorking,
Surrey RH5 6NT \\ 
$^6$ Instituto Nacional de Pesquisas Espaciais/MCT, S.J. dos Campos,
Brazil}

\begin{document} 
\date{Accepted for publication in MNRAS Letters, 2011 August 26}
\pagerange{\pageref{firstpage}--\pageref{lastpage}} \pubyear{2011} 
\maketitle
\label{firstpage}

\begin{abstract} 
The  abundance   ratios  between  key   elements  such  as   iron  and
$\alpha$-process elements  carry a wealth  of information on  the star
formation history (SFH) of galaxies. So far, simple chemical evolution
models  have linked \aFe\  with the  SFH timescale,  correlating large
abundance  ratios  with short-lived  SFH.  The  incorporation of  full
spectral fitting to  the analysis of stellar populations  allows for a
more  quantitative  constraint between  \aFe\  and  the  SFH. In  this
letter,  we provide,  for  the first  time,  an empirical  correlation
between \aFe\ (measured from spectral indices) and the SFH (determined
via a  non-parametric spectral-fitting method). We  offer an empirical
version  of the  iconic  outline of  \citet{Th05,Th10}, relating  star
formation timescale with galaxy mass, although our results suggest, in
contrast, a significant population  of old ($\simgt$10~Gyr) stars even
for   the   lowest   mass   ellipticals  (M$_s\sim   3\times   10^{10}
$M$_\odot$). In addition, the abundance  ratio is found to be strongly
correlated with  the time to  build up the stellar  component, showing
that the highest  \aFe\ ($\simgt +0.2$) are attained  by galaxies with
the   shortest   half-mass  formation   time   ($\simlt  2$~Gyr),   or
equivalently,   with  the   smallest  ($\simlt$   40\%)   fraction  of
populations younger  than 10~Gyr. These  observational results support
the   standard  hypothesis  that   star  formation   incorporates  the
Fe-enriched  interstellar   medium  into  stars,   lowering  the  high
abundance ratio of the old populations.
\end{abstract}

\begin{keywords} galaxies: elliptical and lenticular, cD -- galaxies: evolution -- galaxies:
formation -- galaxies: stellar content -- galaxies: structure. 
\end{keywords}

\section{Introduction}

The  use  of spectral  indices  has  allowed  a relatively  successful
description of the unresolved stellar populations of galaxies in terms
of   three  luminosity-weighted  parameters:   the  age,   the  global
metallicity, and the \aFe\ abundance ratio. The advent of new tools of
analysis  (e.g. spectral  fitting), the  increasing refinement  of the
population   synthesis  models   and  libraries   \citep[e.g.   MILES,
][]{SB06a}   and   the    access   to   vast   spectroscopic   surveys
\citep[e.g.  SDSS,   ][]{sdss}  allows  for  a  fresh   look  at  well
established  standards.  One  of  the current  paradigms involves  the
existence  of a  relationship  between \aFe\  and  the star  formation
timescale.  On the basis of this relationship lies the assumption that
the stellar  abundance ratios  quantify the contribution  of different
{\it  stellar   chronometers}  (e.g.   supernovae)   to  the  chemical
enrichment  of  the subsequent  generations  of stars.   Core-collapse
supernovae, triggered soon after  the onset of star formation, release
mostly  Mg  and   other  $\alpha$-process  elements,  whereas  type~Ia
supernovae, with  a more complex time delay,  contribute large amounts
of    Fe    \citep[see    e.g.][]{gr83,maoz11,man05}.     Hence    the
\aFe\  abundance  ratio is  a  useful  tool  to track  star  formation
timescales $\Delta t\simlt$1Gyr \citep[see e.g.][]{mt87,Th99,fs03}.

By using simple chemical  evolution models, \citet{Th05} established a
theoretical relationship  between the  mass of a  galaxy and  its SFH,
described by a Gaussian function. This Gaussian curve is characterized
by the mean, given by the SSP-equivalent age, and the FWHM, related to
the   \aFe\  abundance   ratios,   such  that   \aFe\   $\sim  0.2   -
0.17\ \log({\rm  FWHM/Gyr})$. Therefore  a high abundance  ratio would
result from  a concentrated burst  of star formation, while  solar, or
sub-solar ratios would reveal a  more extended SFH. The application of
this  rule  to observational  data  by  \citet{Th05} and  \citet{Th10}
produced  an  iconic  outline  \citep[figure 9  in][]{Th10}  in  which
massive  galaxies show  older and  narrower stellar  age distributions
compared to their less massive counterparts.

There are tools currently available to produce more detailed SFHs than
the  Gaussian curves  described  above. The  use  of spectral  fitting
\citep[e.g.][]{CF05} allows us to probe the full range of populations,
giving   a   more  robust   estimate   of  mass-weighted   properties.
Furthermore, our  methodology follows a  non-parametric approach, with
the goal of  establishing a distribution of ages  that better maps the
spectroscopic  data.  These  estimates are  less prone  to  the biases
inherent  in the  use  of luminosity-weighted  properties (when  using
SSPs) or in the parametric assumptions (e.g.  when using exponentially
decaying SFHs).  The analysis  of the abundance ratios follows instead
the traditional method based on  line indices.  We emphasize that both
properties   (SFHs   and  \aFe\   ratios)   are  therefore   extracted
independently.   In the future,  the development  of $\alpha$-enhanced
population   synthesis   models   \citep[e.g.   ][]{Cer07,   Coelho07}  will  open  up  the possibility  of  a  fully-consistent
analysis of the SFHs and  the abundance ratios from a spectral fitting
point of  view~\footnote{ As a robustness  check, we have  derived galaxy
  SFHs  also  with the  ~\citet{Cer07}  $\alpha$-enhanced SSP  models,
  finding  good agreement  with the  results based  on the  widely used,
  solar scaled, MILES models.  }

The goal  of the present study is  to revise the link  between the SFH
and  \aFe\  in  elliptical  galaxies, improving  over  the  insightful
outline of  \citet{Th05,Th10} by using a combination  of {\it spectral
  fitting}, to determine non-parametric  SFHs, and {\it line strength}
analysis to derive abundance ratios.

\section{Sample and Analysis}

Our methodology involves the analysis of the optical spectra to
constrain the properties of the underlying stellar populations. The
sample is extracted from the $\sim$40,000-strong SPIDER catalogue of
early-type galaxies \citep{SP-I} (hereafter Paper I), which covers a
redshift range of 0.05 to 0.095 and absolute magnitudes brighter than
M$_r$ = --20.  The spectroscopic data come from the seventh data
release (DR7) of the Sloan Digital Sky Survey (SDSS) \citep{dr7}.  We
take a subsample with a S/N high enough for a detailed analysis of the
stellar populations. We select a set of 9,999 ETGs, belonging to the
upper quartile of the distribution in S/N, as measured in the {\it
  g}-band (per-pixel S/N$_g >$ 18.3).

Stellar populations are characterized by the following information:
{\bf (I)} mass-weighted average ages ($\langle$age$\rangle_M$) and
metallicities, {\bf (II)} mass-weighted star formation histories
(SFHs) and {\bf (III)} average \aFe\ abundance ratios.  Several
techniques and models are available for the extraction of this
information. In an ongoing study (de la Rosa et al.  in preparation)
the performance of several techniques and models has been evaluated
for SDSS spectra of similar quality. The methodology chosen for the
present study, as described below, results from that performance
comparison.


\begin{figure*}  
\includegraphics[width=174mm]{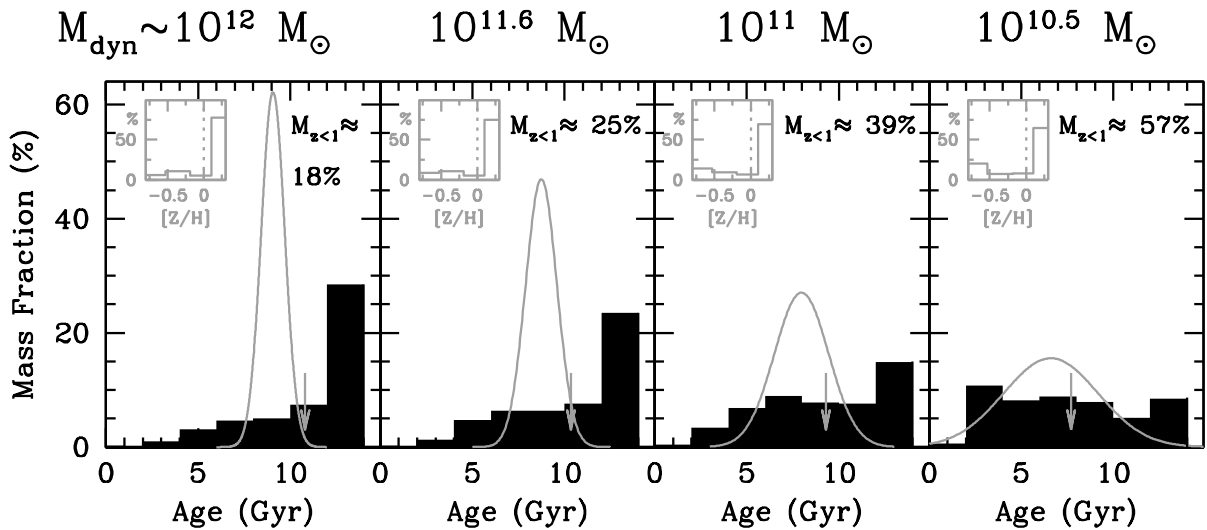}  
\caption{ Variation of the star formation history with respect to
  dynamical mass (each panel representing a mass bin as given in
  table~\ref{tab:binM}).  Sketchy (theoretical) SFHs, plotted as grey
  Gaussian curves, are superimposed over the detailed (empirical) SFH
  histograms (black filled) obtained through spectral fitting.  The
  Gaussian curves are obtained from equation~4 of \citet{Th05}, using
  our SSP-equivalent $\langle$age$\rangle_L$ (+ lookback time) and
  \aFe\ values, with curves normalized to span the same area as the
  histograms. Arrows mark the position of the mass-weighted
  $\langle$age$\rangle_M$, obtained from the SFH histograms. Each
  panel also displays M$_{z<1}$, i.e. the stellar mass percentage
  formed after z=1. The grey insets at the upper-left corners are the
  metallicity histograms, $\langle$[Z/H]$\rangle_M$, for each mass
  bin, with a dotted line at the solar metallicity. }
\label{fig:SFH_M}  
\end{figure*}

\begin{enumerate}

\item[{\bf I)}] Average ages and metallicities are obtained through
  {\it spectral fitting} using the STARLIGHT synthesis code
  \citep{CF05} to find the optimal mixture of single stellar
  populations (SSPs). For the present study, the basis SSPs correspond
  to 76 solar-scaled MILES models \citep{Va10}. These models have a
  Kroupa Universal Initial Mass Function \citep{K01} and span a range
  of 4 metallicities, from Z/Z$_\odot$=1/50 to 1.6, and 19 different
  ages, from 0.5 to 12.6 Gyr.  These models are based on the
  MILES\footnote{\tt www.iac.es/proyecto/miles} stellar library
  \citep{SB06a}, which combines both a dense coverage of the stellar
  atmospheric parameters and a relatively high and nearly constant
  spectral resolution (FWHM=2.3\AA ).  The fitting interval spans from
  3600 to 7350\AA, with emission lines and bad pixels being masked
  out. The extinction due to foreground dust is modelled with the
  CCM-law \citep{CCM89}.  The standard output of the spectral fitting
  consists of the fractional contribution of each SSP to the flux at a
  normalization wavelength $\lambda_0$=4020\AA, converted into a
  stellar mass fraction by the M/L of the given SSP at $\lambda_0$, as
  provided by the models. The detailed population mixture of a galaxy
  can thus be condensed into a single parameter, the mass-weighted
  average age, $\langle$age$\rangle_M$.  The mass-weighted
  metallicity, $\langle[$Z/H$]\rangle_M$, is defined in a similar way.

\item[{\bf II)}] The use of {\it spectral fitting} with solar-scaled
  MILES models provides the fractional contribution of each SSP to the
  total flux, allowing for the reconstruction of the SFH.  Note that
  STARLIGHT does not assume any parametric form, describing the SFH by
  an arbitrary mixture of the basis SSPs. In order to account for the
  range of redshifts of the sample, we add the corresponding lookback
  time to the population ages, ranging from 0.67~Gyr (z=0.05) to
  1.24~Gyr (z=0.095).  Throughout this paper we assume a standard
  $\Lambda$CDM cosmology with $\Omega_m=0.3$ and H$_0$=70 km/s/Mpc.

\item[{\bf III)}] The \aFe\ is measured with the recent models of
  \citet{TMJ11}, which provide absorption-line index values for a
  variety of ages and element abundance ratios. Our procedure fixes
  the age to its SSP-equivalent value, as obtained from spectral
  indices H$\beta$ and [MgFe]'. Then, the Mgb and Fe3 = (Fe4383 +
  Fe5270 + Fe5335)/3 indices are compared to the $\alpha$-enhanced
  model predictions to obtain the (SSP-equivalent) \aFe\ .  To this
  effect, all measured indices are preliminarly corrected for both
  velocity and instrumental dispersion broadening.  The correction
  functions are estimated by convolving a variety of SSP-model spectra
  with broadening functions of different widths.

\end{enumerate}

Dynamical masses are computed with the formula M$_{\rm dyn,n}=
[$K(n)R$_e\sigma^2]/G$, where n is the S\'ersic index and the function
K(n) is taken from \citet{Ber02}. This assumes that mass follows light
in a galaxy, and accounts for structural non-homology of galaxy light
profiles, as parametrized by the Sersic $n$. The three relevant
parameters, n, R$_e$ and $\sigma$, are taken from Paper I.

\begin{table}

\caption{Uncertainties on Parameters}

\label{tab:uncertain} 
\begin{center} 
\begin{tabular}{ccc} 
\hline Parameter & units & typical rms  \\  
\hline SFH age-bars  & percent  & 13.4  \\ 
$\langle$age$\rangle_M$  & Gyrs & 2.7  \\
$\langle[$Z/H$]\rangle_M$   & dex & 0.10  \\ 
\aFe\  & dex & 0.10  \\ 
log(M$_{\rm dyn}$/M$_\odot$)  & dex  & 0.07  \\ 
\hline 
\end{tabular} 
\end{center} 
\end{table}


\begin{table*}

\caption{ Stellar population properties as a funcion of M$_{\rm
    dyn}$. For each mass bin, errors on a given quantity are the
  root-mean-square of deviations from the median value in the bin.}

\label{tab:binM} \begin{center} \begin{tabular}{ccccc} \hline Bin (log M$_{\rm
dyn}$/M$_{\odot}$) & 12.0 &  11.6  & 11.0 & 10.5\\ \hline Log Mass interval &  11.8 -- 12.2
&  11.5 -- 11.7  &  10.9 -- 11.1 & 10.3 -- 10.7 \\ Bin members & 1024 & 1466 & 1408 & 690 \\
$\langle$age$\rangle_M$ (Gyr) & 10.82$\pm$1.68 & 10.35$\pm$1.91 & 9.29$\pm$2.22 &
7.71$\pm$2.45 \\  SSP-equiv. age (Gyr) & 9.07$\pm$2.08 & 8.74$\pm$2.15 & 7.97$\pm$2.47 &
6.61$\pm$2.66 \\  SSP-equiv. \aFe\  & 0.17$\pm$0.09 & 0.15$\pm$0.09 & 0.11$\pm$0.09 &
0.07$\pm$0.10 \\ \hline \end{tabular} \end{center} \end{table*}


\subsection{Error Estimation}

To estimate uncertainties on stellar population parameters, we have
resorted to the 2,283 ETGs in the SPIDER sample with repeated spectral
observations from the SDSS database (see Paper I). After selecting for
pairs of repeated observations with S/N$_g >$18.3 (i.e.  that of the
sample used here), a sub-sample of 393 pairs of spectra is explored
with the same methodology using differences in the extracted
parameters as an estimate of the expected errors.  These uncertainties
are summarized in table~\ref{tab:uncertain}.  It is worth mentioning
that the reported errors refer to single measurements of stellar
population parameters, i.e.  they reflect the statistical uncertainty
inherent in the extraction of SFHs from the noisy spectrum of a given
galaxy. On the other hand, binned quantities in (Figs.~1-2) have much
smaller uncertainties, resulting from average-stacking a large
sub-sample of galaxies in each bin. The comparison of repeated
measurements also shows no systematics in stellar population
parameters as a function of S/N, as shown, for $\sigma_0$, in
\mbox{Paper I.}

\section{Results}

\subsection{The dependence of SFH with M$_{\rm dyn}$} 

Mean SFHs are constructed for four mass bins selected to match the
mass intervals of figure~9 in \citet{Th10}. Our results for these four
mass bins are presented in Table ~\ref{tab:binM}, where one can see
that the age, metallicity, and abundance ratio parameters actually
increase with stellar mass, consistent with previous studies.
Figure~\ref{fig:SFH_M} shows the average distribution of stellar ages
for the ETGs within each mass bin. We emphasize that constraints of
stellar populations from spectroscopic data are more robust when
dealing with {\sl relative} age differences.  By constructing a
stacked age histogram for all galaxies within each mass bin, we give a
rough estimate of the range of stellar ages to be found in galaxies of
the corresponding stellar mass, i.e.  an averaged star formation
history. Note we use seven time intervals only (instead of the
original 19 time steps from the basis SSPs) in order to achieve more
robust results \citep[see discussion in][]{CF05}.  The histograms
combine all possible metallicities within a given age bin.  It is
worth mentioning that the absence of the youngest populations
(t$\simlt 2$~Gyr), evident in the depleted left side of the
histograms, reflects the limit imposed by the lookback time related to
the redshift range of the sample.  We do not consider this a problem
since most of the stellar populations have ages $\simgt 5$~Gyr.  Our
extracted SFHs (filled black histogram in figure~\ref{fig:SFH_M}) are
compared with the ansatz of \citet{Th05,Th10} (grey Gaussian curve),
which relate the mean and variance of a Gaussian SFH with the age and
\aFe\ of the underlying stellar populations. The mean of the Gaussian
SFH is set to our SSP-equivalent age with an added lookback time,
while the FWHM is extracted from the \aFe\ $\sim 0.2 - 0.17\ \log({\rm
  FWHM/Gyr})$ relation given by \citet{Th05}.

We have also constructed stacked metallicity histograms for all
galaxies within each mass bin, displaying them as grey insets in
figure~\ref{fig:SFH_M}. Histograms show high percentages (60--80 \%)
of super-solar metallicities, with a growing contribution (from 6 to
20 \%) of lowest metallicities ([Z/H]$<$-0.5) from the highest to the
lowest galaxy mass bin.  This low metallicity trend might be real, or
just reflect the age--metallicity degeneracy and the fact that
low-[Z/H] stars in the MILES library have super-solar
\aFe\ \citep{MSS11}. To test if this can bias our results, we have
re-computed the age histograms by excluding galaxies with a
significant ($>$ 20\%) low [Z/H]($<$-0.5) population.  We found no
significant variation $wrt$ figure~\ref{fig:SFH_M} (e.g.~ less than
1\% variation in the SFH for age $>$ 10 Gyr).


\begin{figure} 
\includegraphics[width=90mm]{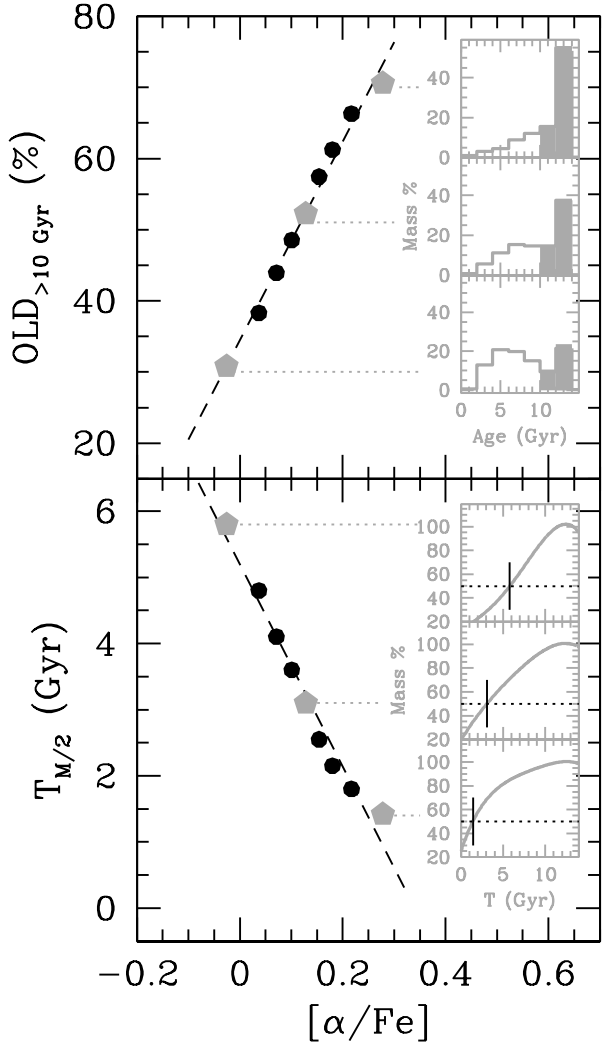} 
\caption{Trends between the SFH and \aFe\ . The top panel shows the
  correlation with the mass fraction in old stars
  ($\simgt$10~Gyr). The insets illustrate individual cases for bins 1,
  5 and 9 of the \aFe\ distribution (grey pentagons). The shaded
  histogram bins in the insets mark the $\simgt $10~Gyr populations
  used for the definition of the OLD parameter. The bottom panel shows
  the trend with half-mass time (i.e. the time interval over which 1/2
  of the total stellar mass is formed). Individual cases of the
  stellar mass growth for bins 1, 5 and 9 are shown in the insets,
  with the black lines guiding the eye for the value at 0.5 of total
  mass.}
\label{fig:OLD} 
\end{figure}


A comparison between our SFH extraction and the Gaussian ansatz of
\citet{Th05,Th10}, shows an overall agreement, although our
non-parametric approach gives a more detailed view of the underlying
SFHs. We emphasize that it is the relative difference across mass bins
what should be compared and not the absolute age values, whose
estimate can be significantly affected by the adopted methodology
(e.g.  use of spectral-fitting vs.  indices) and stellar population
libraries.  Notice also that the center of the Gaussians does not
coincide with that of the SFH histograms, as expected by the fact that
luminosity-weighted ages are systematically younger than their
mass-weighted counterparts \citep[e.g.  ][]{Serr07,Ro10}.  Both the
SFH and the Gaussians show the characteristic trend towards younger
ages and more extended age distributions in lower mass galaxies.

However, figure~\ref{fig:SFH_M} goes beyond a simple description of
the SFH as a Gaussian.  The non-parametric approach of STARLIGHT
allows us to study in more detail the relative change of SFHs with
respect to galaxy mass. One notices that over the mass range probed,
all galaxies feature a significant population of old stars. If we
consider the stellar mass formed at redshifts z$>$1 (i.e. older than
$\sim 8$~Gyr in a local sample) we still get about 40\% of stellar
mass formed even at M$_{\rm dyn}\sim 3\times 10^{10}$M$_\odot$.
Furthemore, the most massive galaxies (M$_{\rm dyn}\sim
10^{12}$M$_\odot$) undergo an intense, short-lived period of star
formation, in which over 50\% of the stellar mass is formed within a
very short period ($\Delta t\simlt 2$~Gyr) at redshifts z$\simgt$4,
followed by a tail of low-level star formation. After z$\sim$1, only
18\% of the stars are newly formed, perhaps the contribution from
minor mergers that could explain the size evolution of massive
galaxies \citep[see e.g.][]{naab09,i3}.

\subsection{The dependence of SFH with \aFe}

In order to confirm a link between the SFH shape with \aFe, we bin the
sample with respect to the abundance ratios into 9 bins, each one
containing 1,111 objects. An average SFH is constructed for each bin
and two parameters are defined to describe the distribution of stellar
ages. The {\it OLD} parameter measures the stellar mass fraction
contributed by the oldest populations, namely above 10 Gyr. The upper
panel of Fig. ~\ref{fig:OLD} shows the increase with \aFe\ of the {\it
  OLD} parameter. A simple linear regression gives for the mass
fraction in old stars:

\be {\rm OLD} = 1.397[\alpha/{\rm Fe}]+0.344.  \ee

\noindent For clarity, three insets show the average SFHs for bins 1,
5 and 9 (corresponding to lowest, intermediate, and highest \,
\aFe). The distribution of the extracted values of the {\it OLD}
parameter for galaxies within the same bin has a standard deviation
around 25\%. The {\it OLD} parameter in these insets corresponds to
the added height of the two shaded oldest bins of the
histograms. According to this result, the drop in the \aFe\ abundance
ratio is motivated by a growing fraction of younger populations on top
of the prevalent old component.  This result supports the standard
interpretation \citep[see e.g.][]{SB06b}, whereby old stellar
populations, formed from primeval ingredients, show high
\aFe\ abundance ratios, further reduced by subsequent star formation
episodes, which incorporate the Fe-enriched inter-stellar material
into stars.

A complementary view of the same result uses the cumulative
integration of the SFH to calculate the ``half-mass time''
(T$_{M/2}$), i.e. the time needed to form 1/2 of the final stellar
mass of the galaxy. We note that the initial time is set to a
lookback-time of 13 Gyr for all galaxies. However, this is an
arbitrary choice depending on the basis of SSP models used for the
{\it spectral fitting}. The lower panel of Fig.~\ref{fig:OLD} shows a
decreasing \aFe\ abundance ratio related to a growing half-mass time
$T_{M/2}$, revealing the connection between an extended period of star
formation and the decrease of the \aFe\ abundance ratio.  Similarly to
the top panel, the three insets show the average cumulative
distribution of the SFH for bins 1, 5 and 9. A simple regression
gives:

\be T_{{\rm M}/2} ({\rm Gyr})= -15.3[\alpha/{\rm Fe}] +5.2,  \ee

\noindent with a scatter around 2.5~Gyr (rms) for galaxies belonging
to the same bin in \aFe.  The fiducial value for solar abundance
ratios is represented, therefore, by a fraction of 34\% in stars older
than 10~Gyr and a half-mass formation time around 5.2~Gyr.

In general, absolute values of stellar population parameters depend
significantly on methods and models used to derive them.  Hence, we
warn the reader that equations (1) and (2) apply exactly only to the
set of models used in the present work, as well as to the lookback
time scheme adopted here to bring all SFHs into a common time-frame.
In other words, using equations~1 and~2 allows one to convert the
\aFe\ measured with \citet{TMJ11} models into half-mass time $T_{M/2}$
and mass fraction in old ($\simgt$10~Gyr) stars, as they would be
measured with solar-scaled MILES models and lookback times added up.

\section{Conclusions}

This letter presents an independent comparison between the
\aFe\ abundance ratios and the star formation history of a
volume-limited sample of $\sim 10^4$ giant early-type galaxies
(M$_r<-20$), extracted from the Sloan Digital Sky Survey.
Figure~\ref{fig:SFH_M} shows the characteristic trend towards more
extended SFHs in lower mass ETGs, in agreement with previous work
\citep[e.g.][]{Th05} but with the additional information regarding the
distribution of stellar ages. We find that all ETGs in our sample have
a significant fraction of very old stars (ages $\simgt $10~Gyr), at
variance with the simplified ansatz of a Gaussian SFH of
\citet{Th05,Th10}. The fraction of stars {\sl formed} after $z\simlt
1$ increases from 18\% at M$_{\rm dyn}\sim 10^{12}$M$_\odot$ up to
57\% for M$_{\rm dyn}\sim 3\times10^{10}$M$_\odot$. This points to the
young populations, formed out of a Fe-enriched medium, as responsible
for the decrease of the initially high \aFe\ abundance ratios of the
old stellar components \citep[see e.g.][]{SB06b}.

In addition to the SFHs, we used individual absorption lines to
determine the \aFe\ abundance ratios, following the prescriptions of
\citet{TMJ11}.  Figure~\ref{fig:OLD} shows the correlation between
these two {\sl independently} estimated properties. We find a strong
correlation between \aFe\ and either the mass fraction in old
($\simgt$10~Gyr) stars ($OLD$ parameter; top panel), or the time to
form one-half of the current stellar mass of the galaxy (T$_{\rm
  M/2}$, bottom). Both cases give a clear picture towards enhanced
\aFe\ for the SFHs that are more short-lived.  Those for which T$_{\rm
  M/2}\simlt 2$~Gyr feature highest, super-solar, abundance ratios,
with \aFe$\simgt0.2$.  Alternatively, an \aFe$ \simgt 0.2$ is reached
by those galaxies where the old populations ($\simgt 10$~Gyr)
constitute more than 60\% of the stellar mass content.  These
quantitative results are powerful calibrators of models of galaxy
formation.

\section*{Acknowledgments}

IGR acknowledges a grant from the Spanish Secretar\'\i a General de
Universidades of the Ministry of Education, in the frame of its
programme to promote the mobility of Spanish researchers to foreign
centers. IF acknowledges a grant from the Royal Society and support
from the IAC to carry out this research project. We have used data
from the SDSS (http://www.sdss.org/collaboration/credits.html).


\label{lastpage} 
\end{document}